\documentclass{ws-procs975x65}

\usepackage{graphicx,epsfig, amsmath,amssymb,amsfonts}

\begin{document}


\title{SUPERNOVAE CONSTRAINTS ON COSMOLOGICAL DENSITY PARAMETERS AND COSMIC 
TOPOLOGY\footnote{This research has been partially supported by CNPq.}} 

\vspace{-1mm}
\author{MARCELO J. REBOU\c{C}AS}
\address{Centro Brasileiro de Pesquisas F\'{\i}sicas\\
Rua Dr.\ Xavier Sigaud 150, \  22290-180 Rio de Janeiro --
RJ, Brazil}

\begin{abstract}
We illustrate the constraints that a possible detection of a non-trivial 
spatial topology may place on the cosmological density parameters by 
considering the $\Lambda$CDM model Poincar\'e dodecahedal space 
(PDS) topology as a circles-in-the-sky detectable topology. 
To this end we reanalyze the type Ia supernovae constraints on the 
density parameter plane $\Omega_k$~--~$\Omega_{\Lambda}$ and show that 
a circles-in-the-sky detectable PDS topology gives rise to important 
constraints on this parameters plane.
\end{abstract}

\bodymatter

\section{Introduction}\label{intro}

Let us begin by stating our basic context. In the light of the current 
observations, we assume that the current matter content of the Universe 
is well approximated by a dust of density $\rho_m\,$, (baryonic plus dark 
mater) plus a cosmological constant $\Lambda$ ($p = -\rho_{\Lambda}$). 
In this $\Lambda$CDM context, we additionally assume that the Universe 
is modelled by a space-time manifold 
$\mathcal{M}_4 = \mathbb{R}\times M_3$ with a locally (spatially) 
homogeneous and isotropic  metric
\begin{equation}
\label{RWmetric}
ds^2 = - dt^2 + a^2 (t) \left [ \frac{dr^2}{1-kr^2} + r^2 (d\theta^2
+ \sin^2 \theta \, d\phi^2) \right ] \,,
\end{equation}
where, depending on the spatial curvature $k$,
the geometry of the $3$--space $M_3$ is either Euclidean ($k=0$),
spherical ($k=1$), or hyperbolic ($k=-1$). 

The metric~(\ref{RWmetric}) only expresses the principle of spatial 
homogeneity and isotropy along with the existence of a cosmic time $t$, 
it does not specify the underlying space-time manifold $\mathcal{M}_4$
nor the spatial section $M$, which are often taken to be the 
(simply-connected) spaces: Euclidean $\mathbb{E}^3$, spherical 
$\mathbb{S}^3$, or hyperbolic space $\mathbb{H}^3$.
This has led to a common misconception that the Gaussian curvature 
$k$ of $M_3$ is all one needs to determine the topology of the 
spatial sections $M_3$.  However, it is a mathematical result that
the great majority of constant curvature $3$--spaces, $M_3$,  are 
multiply-connected quotient manifolds of the form $\mathbb{R}^3/\Gamma$, $\mathbb{S}^3/\Gamma$, and $\mathbb{H}^3/\Gamma$, where $\Gamma$ is a 
fixed-point free group of isometries of the corresponding covering space
(see Ref.~\refcite{CosmTopReviews} for details).

In the general relativity (GR) framework 
the Friedmann written in the form $\Omega_k= \,\Omega_{m}+ \,\Omega_{\Lambda}-1\,$ 
makes apparent that the chief point in
the search for the spatial curvature (and the associated geometry) 
is to constrain the total density 
$\Omega_{\mathrm{tot}}=\,\Omega_{m}+ \,\Omega_{\Lambda}$ from 
observations. This amounts to determining regions in the 
parametric planes $\Omega_{m}\,$--$\,\,\Omega_{\Lambda}$ (or 
$\Omega_{m}\,$--$\,\,\Omega_{k}$), which consistently account for 
the observations, and from which one expects to deduce the geometry 
of the Universe.

Now given that the spatial geometry can be probed but its 
knowledge does does not determine the topology of the $3$--space $M_3$, 
the question arises as to whether the spatial topology is an observable 
that can be be used to set constraints on the density parameters. 
$\Omega_m$ and $\Omega_{k}$. 
An important observational consequence of a observable 
nontrivial topology\cite{TopDetec} of $M_3$ is the existence
of the circles-in-the-sky,\cite{CSS1998} i.e., pairs of 
correlated circles with the same fluctuation of temperature 
distribution $\delta T$ will be imprinted on the CMBR anisotropy 
sky maps~\cite{CSS1998,CGMR05}. Hence, to observationally probe 
a putative a nontrivial topology of $M_3$, one ought to examine 
the full-sky CMBR maps in order to extract the pairs of correlated 
circles and determine the spatial topology.

Here we briefly illustrate the constraints that a possible detection 
of a non-trivial spatial topology may place on the cosmological 
density parameters in the $\Lambda$CDM, i.e.,  by assuming a Poincar\'e 
dodecahedral (PDS) topology as a circles-in-the-sky observable spatial 
topology we reanalyze the type Ia supernovae constraints on the 
density parameter plane $\Omega_k$~--~$\Omega_{m}$ that arise from 
{\em gold} set of 157 SNe Ia, as compiled by Riess \emph{et al.\/}~\cite{Riess2004}, 
and show that a circles-in-the-sky detection of the PDS topology gives rise to 
important additional constraints on  parameters of this plane.

\section{Constraints and Concluding Remarks} \label{Constr-Concl} 

Using the three-year data the WMAP team\cite{WMAP-Spergel:2006} reports six 
different values for the total density $\Omega_{\mathrm{tot}}$ ranging 
from a very nearly flat $\Omega_{\mathrm{tot}}= 1.003^{+0.017}_{-0.013}$ 
to positively curved $\Omega_{\mathrm{tot}}= 1.037^{+0.015}_{-0.021}$ 
depending on the combination of data set used to resolve the geometrical 
degeneracy.
The Poincar\'e dodecahedral space (PDS), $\mathcal{D}=\mathbb{S}^3/I^\star$,
fits both this latter density and the suppression of power of the low 
multipoles observed by the WMAP team.%
\footnote{Other spherical topologies, notably $\mathcal{O}=\mathbb{S}^3/O^\star$, 
where $O*$ is the binary octahedral group, also explain these data 
analysis.\cite{Aurich12}}
Attempts to find antipodal or nearly-antipodal circles-in-the-sky in the
WMAP data have failed~\cite{Cornish,Key-et-al:2006}. On the other hand,
hints of matching circles\cite{Roukema} in first year ILC WMAP 
maps have be found, but although a second group has confirmed 
the circles,\cite{Key-et-al:2006} they have also shown that 
the circle detection lies below the false positive threshold.\cite{Key-et-al:2006}. 
Even if one adopts  the result that pairs of antipodal (or nearly antipodal) 
circles of radius $\gamma \geq 5^\circ$ are undetectable in the current CMBR maps
the question arises as to whether the circles are 
not there or are simply hidden or destroyed by various sources of contamination
or even due to the angular resolution of the current CMBR 
maps.\cite{Holger2006} 
The answer to these questions requires 
great care, among other things, because the level 
of contamination depends on both the choice of the cosmological models 
(parameters) and on the topology\cite{Weeks2006}. Results so far remain
non-conclusive, i.e., one group finds their negative outcome to be robust 
for globally homogeneous topologies, including the dodecahedral space, in 
spite of contamination\cite{Key-et-al:2006}, while another group 
finds the contamination strong enough to hide the possible 
correlated circles in the current CMBR maps.\cite{Aurich2006,Holger2006}
Thus, it is conceivable that the correlated circles may have been overlooked 
in the CMBR sky maps searches.
\begin{figure}[!htb]
\begin{center} 
\includegraphics[width=4.5cm,height=4.50cm,angle=0]{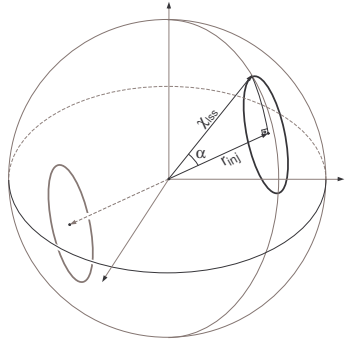}
\caption{\label{CinTheSky1} A schematic illustration of two antipodal
matching circles in the LSS.} 
\end{center}
\end{figure}

In the Poincar\'e dodecahedral space $\mathcal{D}$, which we shall assume
here,  the pairs of matching circles are necessarily 
antipodal as shown in Fig.~\ref{CinTheSky1}.
Clearly the distance between the centers of each pair of the 
correlated  circles is twice the injectivity radius of the smallest sphere 
inscribable  $\mathcal{D}$.
A straightforward use of trigonometric relations for the right-angled 
spherical triangle shown in Fig.~\ref{CinTheSky1} yields 
\begin{equation}
\label{ChiLSS}
\chi^{}_{lss}= \frac{d^{}_{lss}}{a_0}\! =\! 
\sqrt{|\Omega_k|} \int_1^{1+z_{lss}} \hspace{-7mm}
\frac{dx}{\sqrt{\Omega_{m}x^3\! + \Omega_k x^2 
\!-(\Omega_{m}\!+\! \Omega_{k})\!+\!1 }}= \tan^{-1} 
\left[\,\frac{\tan r_{inj}}{\cos \alpha}\, \right],
\end{equation}
where $d^{}_{lss}$ is the radius of the LSS, $x=1+z$ is an integration
variable, $\Omega_k =1-\Omega_{\mathrm{tot}}$, $r_{inj}$ is a 
topological invariant that is equal to $\pi/10$ for $\mathcal{D}$, 
the distance $\chi^{}_{lss}$ is measured \emph{in units of the
curvature radius}, 
$a_0=a(t_0)=(\,H_0\sqrt{|1-\Omega_{\mathrm{tot}}|}\,)^{-1}\,$,
and $z_{lss}=1089$~\cite{WMAP-Spergel:2003}.
Equation~(\ref{ChiLSS}) makes apparent that $\chi^{}_{lss}$ depends on
the cosmological scenario, which we have taken to be $\Lambda$CDM. 

Equation~(\ref{ChiLSS}) give the relation between 
the angular radius $\alpha$ and the parameters of the $\Lambda$CDM 
model, and thus can be used to set constraints on these parameters. 
For a detailed analysis of topological constraints in the context 
of other models and data sets, including braneworld inspired models,  
we refer the readers to  
Refs.~\refcite{Previous1}  and Refs.~\refcite{Previous2}.

\begin{figure*}[ht!]
\begin{center}
\includegraphics[height=4.9cm]{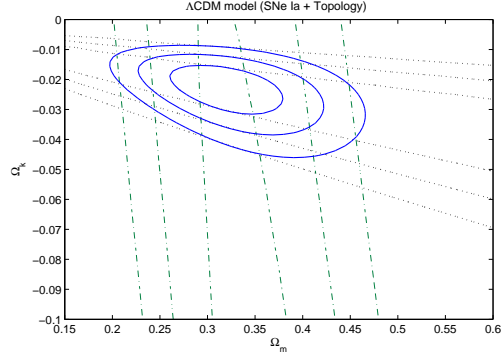} 
\caption{\label{LCDM_p0}Confidence regions ($1$-$\sigma$, $2$-$\sigma$  
and $3$-$\sigma$) 
in the plane $\Omega_m-\Omega_k$ $\Lambda$CDM model obtained 
with the SNe Ia gold sample assuming a ${\cal D}$ space topology with 
$\gamma=50^o \pm 6^\circ$. Also shown are
the contours obtained assuming no topological data (dash-dotted
lines) and the ones corresponding to topology only (dotted lines).}
\end{center}
\end{figure*} 

To illustrate the role of the cosmic topology in constraining the 
density parameter in the context of $\Lambda$CDM model, we consider 
the $\mathcal{D}$ spatial topology, and assume  
the angular radius $\alpha = 50^\circ$ and uncertainty 
$\delta {\alpha} \simeq 6^\circ$.  
Figure~\ref{LCDM_p0} shows the results of our joint SNe Ia gold sample%
\cite{Riess2004} plus cosmic topology analysis. There we display the 
confidence regions in the parametric plane $\Omega_{k}\,$--$\,\,\Omega_{m}$ 
and also the regions from the conventional analysis.
The comparison between these regions makes clear that
the effect of the $\mathcal{D}$ topology is to reduce considerably 
the area corresponding to the confidence intervals in the parametric 
plane as well as to break degeneracies arising from the current SNe 
Ia measurements. The best-fit parameters for this joint analysis are
$\Omega_m = 0.32$ and $\Omega_k =-0.022\,$.

\section*{Acknowledgments} \vspace{-1mm}
The author is grateful to J.S. Alcaniz, M.C. Bento, O. Bertolami, 
M. Makler, B. Mota, N.M.C. Santos and P.T. Silva for valuable 
discussions. 


\end{document}